# A generalized 3D elastic model for nanoscale, self-assembled oxide-metal thin films with pillar-in-matrix configurations

Kyle Starkey, Ahmad Ahmad, Juanjuan Lu, Haiyan Wang, Anter El-Azab*

*School of Materials Engineering, Purdue University, West Lafayette, IN 47907, USA*

*Corresponding author: Anter A. El-Azab, aelazab@purdue.edu

**Abstract:**

In recent years, functional oxide-metal based vertically aligned nanocomposite (VAN) thin films have gained interest due to their intriguing physical properties and multifunctionalities stemming from the complex interactions between the two phases in the film and the substrate. In this work, we develop a model for studying the energetics of these thin film systems, including the effects of both lattice mismatch and capillary forces due to interface curvature. Each phase is incorporated into the model using a phase indicator function, and we introduce the capillary forces as body forces using a vector density representation of the interface. The model is implemented using the finite element method to study the deformation of the thin film which is composed of Au nanopillars embedded in a $La_{0.7}Sr_{0.3}MnO_3$ (LSMO) matrix on an $SrTiO_3$ (STO) substrate. The results suggest that the total energy is lowest for random configurations of pillars compared to ordered square and hexagonal lattice configurations, consistent with the random distribution of pillars found in experiments. Furthermore, we find that the interfacial energy dominates the total energy of each configuration, suggesting that interfacial energy in the system is an important design parameter for nanocomposite growth, along with the lattice mismatch.

**1. Introduction**

Firstly discovered in 2002, vertically aligned nanocomposites (VANs) refer to novel nanosized composites with a structure of vertically aligned nanopillars or columns embedded inside the matrix [1]. Until now, known systems include oxide-oxide [2–4], oxide-metal [5,6], nitride-nitride [7], oxide-nitride [8], nitride-metal [9] VANs, and more combinations such as complex core-shell heterostructures [10], nitride-oxide-metal [11] and oxide-metal-oxide [12] three-phase VANs. Research into oxide-metal VANs is emerging rapidly, primarily due to their appealing potential in the combinations of optical, electrical, magnetic, ferroelectric, and electro transport properties. Future device integrations of oxide-metal VANs are more attractive: metamaterials for optical applications such as hyper lenses [13], memory devices for data storage purposes [14], electrolyte materials for lithium batteries and solid fuel cells [15], to name a few.

Though there are growing reports on novel oxide-metal VAN systems, material selection and precise control of VAN morphology remain uncertain experimentally. On the one hand, the heteroepitaxial growth of two phases inevitably introduces strain via lattice coupling on the large interfacial areas in both lateral and vertical directions [16]. On the other hand, the growth of either nanopillars or matrix phases also depends on the interfacial energies [16]. The interplay between these two factors, mismatch strain and interfacial energies, decides the final morphology and ordering of a VAN system considering it reaches a thermodynamical equilibrium [17]. Such systems, however, are seldom in global equilibrium states due to kinetic limitations. The states reached during growth are thus mostly metastable equilibrium states, and the investigation of their energetics offers insight into the experimental observations even without fully investigating the growth kinetics. To analyze the crucial role played by strain inside the film, a strain compensation model has been previously proposed and optimized according to the oxide-metal VAN systems [18], [19]. Another experimental example of exploration into the growth of films is the construction of various 3D oxide-metal heterostructures [20]. A systematic energy calculation and simulation for oxide-metal VANs is in great need.

Despite the extensive experimental demonstrations on VANs in the literature, the growth of such nanocomposites has scarcely been studied thermodynamically in the context of self-assembly of complex oxide systems, although the thermodynamics play a vital role in phase segregation during phase growth [21]. For a particular configuration in the VAN system, the energy can be determined from two components: interfacial and elastic strain energy. The interfacial energy is an estimation of the energy needed to keep two phases bonded [22]. On the other hand, in a multiphase system, the elastic strain energy arises due to the lattice and thermal mismatch of the dissimilar materials. This mismatch gives rise to residual stresses in the material [23]. Consequently, the energy of the VAN systems includes a mechanical component that can be formulated from the theory of elasticity [24, 25]. This mechanical portion of the energy also allows for strain engineering along the interfaces due to the large vertical interface areas [26].

In VAN systems, atoms near the interface have a different environment, and in turn energies, compared to those in the interior. The difference in atomic energies between interface and bulk atoms results in excess free energy known as the interfacial free energy [27]. Because of this energetic difference a force exist at the interface called surface tension [28–30]. Moreover, in curved solid surfaces this capillarity effect is balanced by the jump in traction across the interface [29, pp. 71–73]. Leo and Sekerka modeled the crystal-crystal interface surface stress as a capillary force that induces a strain field in the bulk by considering the surface free energy density, and the curvature of the interface in the reference state [31]. Bico et al studied the elastocapillary phenomena by examining the deformation process of stiff solid bulk material in contact with fluid drop due to the presence of capillary forces [32].

There have been numerous studies on the thermodynamics of solid interfaces [24, 26, 31, 34–36]. Other studies highlight the effects of surface stresses which have an impactful role in the phase equilibria [32, 37–39]. These studies highlight the importance of the thermodynamics of interfacial quantities such as surface stress that can drive certain instabilities toward achieving a phase transformation equilibrium. In another study, the impact of interfacial stresses on an effective elastic moduli of nanocomposite materials was examined by formulating a model that considers the effect of interfacial stress [39]. A recent interesting study that takes into account the curvature effects on the interfacial energy by studying the influence of surface stresses on spherical nanoparticle elastic material properties has been conducted [40]. In another study by Gurtin, the interface deformation kinematics was highlighted by the capillary effects at the interface that is correlated with the interfacial curvature [41].

In this work, an elastic model is developed to characterize the lattice mismatch deformation of a 3D epitaxial two-phase thin film on a dissimilar substrate. The capillarity effects at the interface are represented by concentrated body forces in the elastic model. A boundary value problem for the crystal equilibrium equations is formed and its solution is obtained using the finite element method. We organize the paper as follows. A brief introduction to the experimental observation of VAN thin films is introduced in Section 2. Then, section 3 analyzes the kinematics, interface representation, and a derivation of the governing equations. Following this, the weak form of the governing equations is introduced. Next, we introduce a representative area element (RAE) in Section 5 to compare the energy of various pillar morphologies consistently. Lastly, in Section 6, we analyze the energy of various pillar morphologies followed by concluding remarks and a brief discussion of future work.

**2. The VAN system and its idealization**

The La$_{0.7}$Sr$_{0.3}$MnO$_3$-Au thin films reported by Lu *et al.* [51] is a typical class of VAN systems to which the current work is directed. This system was synthesized using the pulsed laser deposition (PLD) technique onto 5mm×10mm×0.5mm SrTiO$_3$ (STO) substrates in a 50mTorr oxygen background, with a laser energy of 420mJ and deposition frequency of 5Hz. Film morphology was revealed by transmission electron microscopy (TEM), which is depicted in Fig. 1. Two cross-sectional images and one plane view image were combined into a 3D illustration. In this film, the Au nanopillars were treated as perfect cylinders with an average radius of 5.64nm, and a volume ratio of 15%. The film thickness is 84.37nm. Fig. 1d is a model showing an ideally ordered distribution of Au nanopillars embedded in La$_{0.7}$Sr$_{0.3}$MnO$_3$(LSMO) matrix. In the sequel, this configuration will be considered for mechanics modeling keeping or perturbing the order.

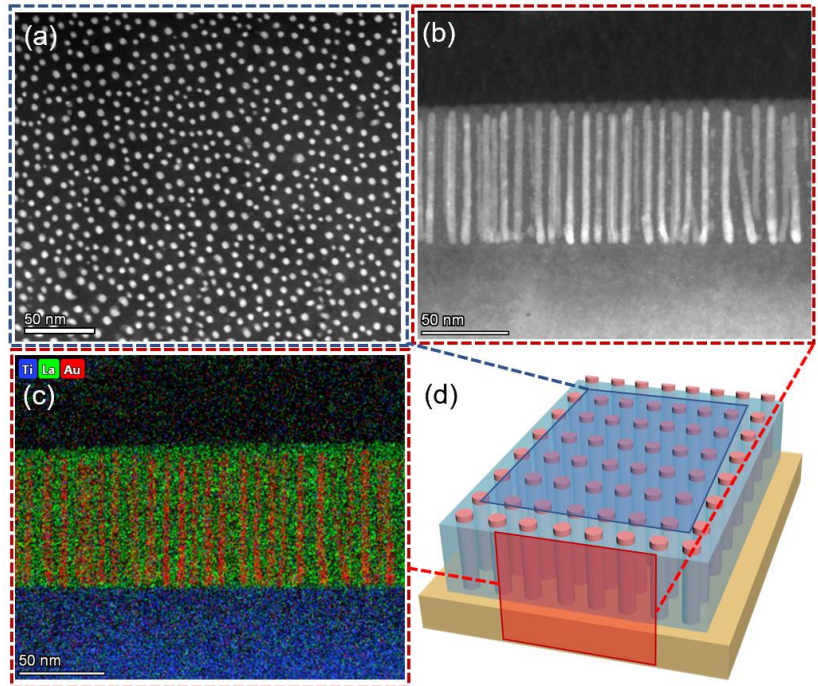

Fig. 1: (a) A plan-view STEM image; (b) HADDF cross section image; (c) EDS-mapping of the cross section of the film; (d) A 3D visualization of idealized LSMO-Au VAN structure.

## 3. Modeling of the mechanics of the VAN system with capillary effects

### 3.1. Notation and preliminary content

Here we follow the tensor notation and convention used by Gurtin [42]. In the following equations, we define various differential operators, and in doing so **T** represents a second order tensor field, **v** represents a vector field, and $\phi$ a scalar field

$$\text{grad}(\phi) = \frac{\partial \phi}{\partial x_i} \tag{1}$$

$$\text{div}(\mathbf{v}) = \frac{\partial v_i}{\partial x_i} \tag{2}$$

$$\text{curl}(\mathbf{v})_i = \epsilon_{ijk}\frac{\partial v_k}{\partial x_j} \tag{3}$$

$$\text{div}(\mathbf{T})_i = \frac{\partial T_{ij}}{\partial x_j} \tag{4}$$

The spatial differential operators concerning referential coordinate **X** are denoted by upper case; for example, the divergence of a vector field $\mathbf{v}(\mathbf{X})$ is

$$\text{Div}(\mathbf{v}) = \frac{\partial v_i}{\partial X_i}. \tag{5}$$

We also use the following tensor identities for the double inner product of two second-order tensors **A** and **B**:

$$\mathbf{A}:\mathbf{B} = A_{ij}B_{ij}, \tag{6}$$

We denote surface derivatives with a subscript $\Gamma$. For instance, $\text{div}_\Gamma(\cdot)$ denotes the surface divergence. We refer the reader to [33] for more information on surface derivatives and surface kinematics.

## 3.2. Kinematics

The static deformation of the various phases in the VAN system configuration is described by the deformation mapping, $\mathbf{x} = \vartheta(\mathbf{X})$ with $\mathbf{X}$ and $\mathbf{x}$ being the position vectors in the reference and deformed configurations of the crystal, respectively. In the finite deformation setting, the multiplicative decomposition of the deformation gradient is used to represent the elastic distortion, denoted by $\mathbf{F}^e$, and inelastic distortions, denoted by $\mathbf{F}^v$, in the body and is given by

$$\mathbf{F} = \frac{\partial \mathbf{x}}{\partial \mathbf{X}} = \mathbf{F}^e \mathbf{F}^v. \tag{7}$$

In this expression, the elastic and inelastic distortions are generally incompatible and thus do not globally represent gradients of a deformation mapping. The inelastic distortion accounts for the homogenous stress-free deformation of the body, measured from a reference lattice. It is generally due to lattice mismatch and thermal mismatch among the VAN phases. The elastic distortion accommodates this inelastic distortion in the form of a coherency strain in each phase. We depict this accommodation in Fig. 2. The inelastic distortion maps vectors from the reference configuration to vectors in the intermediate space. The elastic distortion then takes vectors in the intermediate space and maps them to the deformed configuration.

To derive the inelastic distortion, $\mathbf{F}^v$, a reference lattice consisting of a repeating reference unit cell is introduced. In this reference lattice, the pillar, substrate and matrix phases are denoted by $\Omega^p, \Omega^s$ and $\Omega^m$, respectively, and collectively by $\Omega^\alpha, \alpha \in [p, s, m]$, where $\Omega^\alpha \cap \Omega^\beta = \emptyset, \alpha \neq \beta$.

We apply a linear mapping $\boldsymbol{\psi}^\alpha$, $\alpha \in [p, s, m]$, over each phase to deform the reference lattice into the lattices corresponding to each phase in the intermediate space. The homogeneous deformation is shown in Fig. 2.

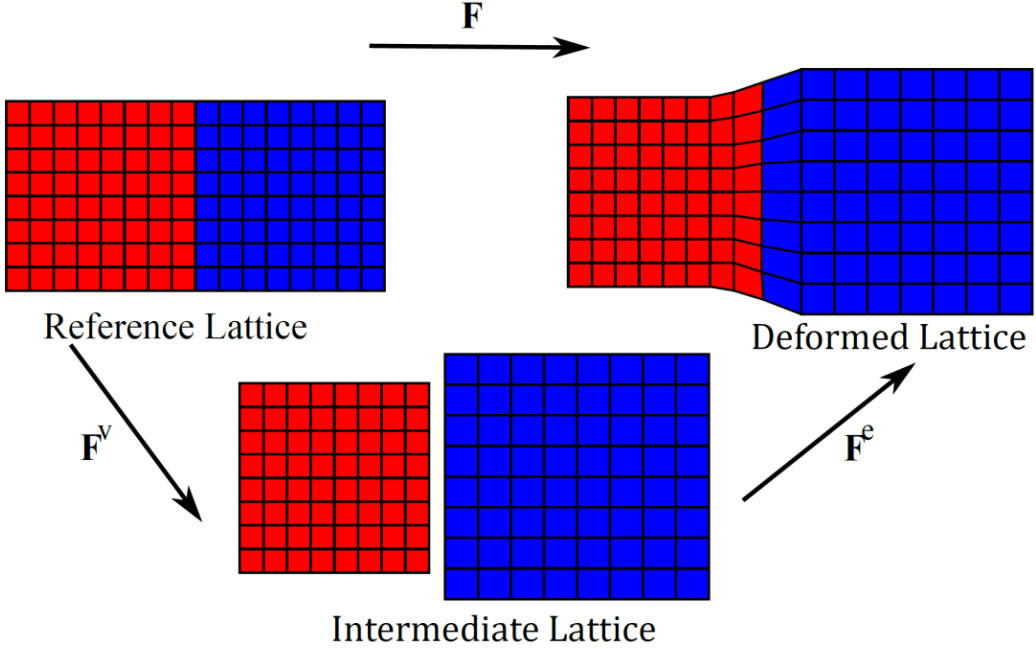

Fig. 2: Stress-free strain applied to each phase in the body resulting in a transformation of the reference lattice to each phase's unit cell. The application of this strain results in a segmented intermediate configuration which is stitched back together with the elastic distortion.

We write the homogenous strain in the form

$$\boldsymbol{\psi}^\alpha = \mathbf{F}_{\Omega^\alpha}\mathbf{X}, \quad \mathbf{X} \in \Omega^\alpha \tag{8}$$

$$\mathbf{F}_{\Omega^\alpha} = \begin{bmatrix} a^x_{\Omega^\alpha} & 0 & 0 \\ 0 & a^y_{\Omega^\alpha} & 0 \\ 0 & 0 & a^z_{\Omega^\alpha} \end{bmatrix}.$$

In the last expression, $a^x_{\Omega^\alpha}$, $a^y_{\Omega^\alpha}$, and $a^z_{\Omega^\alpha}$ are the lattice constants of the unit cells in the **x**, **y**, and **z**-direction, respectively, for the phase $\Omega^\alpha$. If, in addition to deformation, there should be a rotation of the reference lattice through the transformation process $\boldsymbol{\psi}^\alpha$, then this rotation also needs to be accounted for depending on the different orientation relationships between each phase. However, the epitaxy is Au(002)||LSMO(002), Au(002)||STO(002), and LSMO(002)||STO(002) as reported in [43], and thus we omit the lattice rotation in (8). In our model, we assume an fcc-unit cell for Au with lattice parameter, $a^x_{\Omega^p} = a^y_{\Omega^p} = a^z_{\Omega^p} = 4.080\,\text{Å}$, and sc-unit cells for both LSMO and STO with $a^x_{\Omega^m} = a^y_{\Omega^m} = a^z_{\Omega^m} = 3.875\,\text{Å}$ and $a^x_{\Omega^s} = a^y_{\Omega^s} = a^z_{\Omega^s} = 3.905\,\text{Å}$, respectively. Taking $\Omega^j$ the phase to be our reference lattice, the inelastic distortion is given by

$$\mathbf{F}^v = \begin{cases} \mathbf{F}_{\Omega^i}\mathbf{F}_{\Omega^j}^{-1}, & \mathbf{X} \in \Omega^\alpha \\ \mathbf{I}, & \mathbf{X} \in \Omega^\alpha \end{cases} \tag{9}$$

To study the elastic energy of the system, we need to introduce the elastic strain. For this we decompose the Lagrangian strain into elastic and inelastic parts using the multiplicative decomposition

$$\mathbf{E} = \mathbf{F}^{vT}\mathbf{E}^e\mathbf{F}^v + \mathbf{E}^v, \tag{10}$$

where

$$\mathbf{E}^v = \frac{1}{2}(\mathbf{F}^{vT}\mathbf{F}^v - \mathbf{I}) \tag{11}$$

$$\mathbf{E}^e = \frac{1}{2}(\mathbf{F}^{eT}\mathbf{F}^e - \mathbf{I}).$$

The multiplicative decomposition also introduces a variety of stress measures depending on the configuration of their respective force and area components. We are particularly interested in the Cauchy stress, denoted by $\boldsymbol{\sigma}$ and the second Piola-Kirchoff (PK2) stress in the intermediate configuration, which is denote by $\mathbf{S}^e$. The two stress measures are related by

$$\mathbf{S}^e = J^e\mathbf{F}^{e-1}\boldsymbol{\sigma}\mathbf{F}^{e-T}. \tag{12}$$

With this relationship, $\mathbf{S}^e$ can be viewed as the pull back of the Cauchy stress onto the intermediate configuration. The stored elastic energy density in the intermediate configuration is given by

$$W = \frac{1}{2}\mathbf{S}^e : \mathbf{E}^e. \tag{13}$$

### 3.3 Equilibrium equations at the interface and common line

The interface between phase $\Omega^\alpha$ and $\Omega^\beta$ is denoted by $\Gamma^{\alpha\beta} = \overline{\Omega}^\alpha \cap \overline{\Omega}^\beta$ where the overbar denotes the set theoretic closure and the union of all interfaces by $\Gamma = \bigcup_{\alpha,\beta\ \beta>\alpha}^{n} \Gamma^{\alpha\beta}$. We note that the phases $\Omega^p$ and $\Omega^m$ need not be simply connected, and so $\Gamma$ is in general not simply connected. For thin films $\Omega^p$ will, for example, comprises the volume of multiple pillars. The intersection of three different interfaces $\Omega^\alpha$, $\Omega^\beta$ and $\Omega^\gamma$ forms a common line denoted by $C^{\alpha\beta\gamma}$ and the union of all common lines by $C$. The whole body, denoted by $\Omega$, is then $\Omega = (\bigcup_{\alpha=1}^{n} \Omega^\alpha) \cup \Gamma \cup C$ the union of all the phases, interfaces, and common lines. We follow [33] to derive the bulk, interface, and common line momentum balance laws. First, a control volume is introduced within the body with boundary, S. The control volume is placed in an arbitrary region containing interfaces and a common line. This is depicted in Fig. 3.

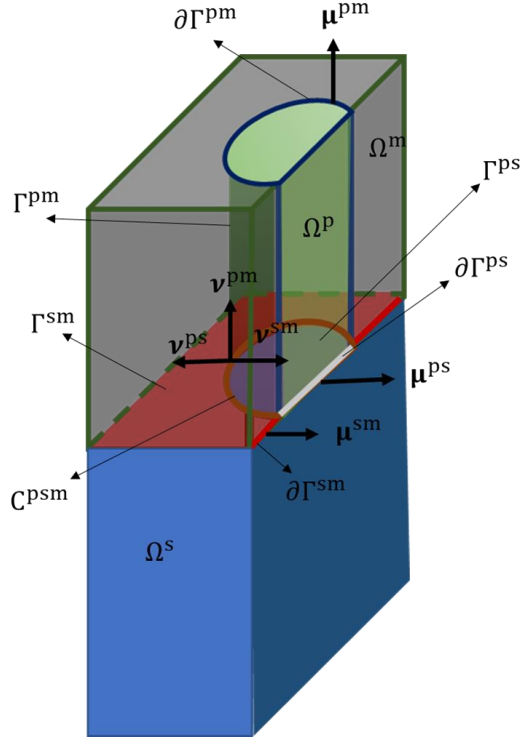

Fig. 3: A control volume $\Omega$ with boundary S containing three interfaces $\Gamma^{ps}, \Gamma^{pm}$ and $\Gamma^{sm}$ which intersect S along the curves $\partial\Gamma^{ps}, \partial\Gamma^{pm}$ and $\partial\Gamma^{sm}$ respectively. The outward-pointing normal to these curves, which also lie on the interface, is denoted by $\boldsymbol{\mu}^{ps}, \boldsymbol{\mu}^{pm}$ and $\boldsymbol{\mu}^{sm}$ respectively. These interfaces also share a common line denoted by $C^{psm}$. The inward-pointing normal to the common line, which lies on each interface, is denoted by $\boldsymbol{\nu}^{ps}, \boldsymbol{\nu}^{pm}$ and $\boldsymbol{\nu}^{sm}$ respectively.

Next, for the quasi-static case, the momentum balance law is given as

$$\mathbf{0} = \int_S \boldsymbol{\sigma} \cdot \mathbf{n}\, da + \int_{\partial\Gamma} \boldsymbol{\sigma}^{(\Gamma)} \cdot \boldsymbol{\mu}\, ds. \tag{14}$$

In this expression, $\mathbf{n}$ is the outward pointing unit normal to the control volume, and $\boldsymbol{\mu}$ is the unit vector that lies on the interface and points out of the control volume on the line created by the intersection of the interface and the control volume boundary. The first and second term in (14) are rewritten using the divergence theorem in the presence of an interface and the surface divergence theorem in the presence of a common line [33],

$$\int_S \boldsymbol{\sigma} \cdot \mathbf{n} \, da = \int_R \text{div}(\boldsymbol{\sigma}) dv + \int_\Gamma [\boldsymbol{\sigma}] \cdot \mathbf{n} \, da \qquad (15)$$

$$\int_{\partial \Gamma} \boldsymbol{\sigma}^{(\Gamma)} \cdot \boldsymbol{\mu} \, ds = \int_\Gamma \text{div}_\Gamma(\boldsymbol{\sigma}^{(\Gamma)}) da + \int_C \left(\left(\boldsymbol{\sigma}^{(\Gamma)} \cdot \mathbf{v}\right)\right) ds$$

In the last expression, dv, da and ds are differential volume, area, and line elements respectively. In the last expression [·] denotes the jump of the quantity in the brackets across the interface and $\left(\left(\cdot\right)\right)$ is defined as

$$\left(\left(\phi^\Gamma \mathbf{v}\right)\right) \equiv \phi^{\Gamma,ps} \mathbf{v}^{ps} + \phi^{\Gamma,pm} \mathbf{v}^{pm} + \phi^{\Gamma,sm} \mathbf{v}^{sm} \qquad (16)$$

where $\mathbf{v}^{ps}$ is the inward-pointing normal to the common line lying on the pillar/substrate interface and $\phi^{\Gamma,ps}$ denotes the limiting value of the field $\phi^\Gamma$ on the pillar/substrate interface connected to the common line. Inserting (15) into (14) and for an arbitrary control volume gives,

$$\text{div}(\boldsymbol{\sigma}) = \mathbf{0} \text{ in } \cup_{\alpha=1}^n \Omega^\alpha \qquad (17)$$

$$[\boldsymbol{\sigma}] \cdot \mathbf{n} + \text{div}_\Gamma(\boldsymbol{\sigma}^{(\Gamma)}) = \mathbf{0} \text{ on } \Gamma \qquad (18)$$

$$\left(\left(\boldsymbol{\sigma}^{(\Gamma)} \cdot \mathbf{v}\right)\right) = \mathbf{0} \text{ on C.} \qquad (19)$$

Equations (17),(18), and (19) represent the bulk, interface, and common line momentum balance, respectively.

**3.4 Interface representation**

In the following sections, it will be convenient to extend each phase to the whole domain Ω so that we can rewrite integrals over each phase as integrals over the whole domain. To do this, the phase indicator function or the volume fraction is introduced, denoted by χ and defined by

$$\int_\Omega \chi^\alpha f dV = \int_{\Omega^\alpha} f dV \tag{20}$$

where $\Omega^\alpha$ is the volume that phase $\chi^\alpha$ represents and f is some arbitrary function defined over Ω. The phase indicator function takes on a value of 1 in the material region $\Omega^\alpha$ and a value of 0 outside. In this sense $\chi^\alpha$ demarcates regions in Ω occupied by the volume $\Omega^\alpha$ and thus we sometimes refer to $\chi^\alpha$ as the volume it represents. We are interested in constructing this type of representation for the pillar, matrix, and substrate phases which are denoted by $\chi^p, \chi^m$ and $\chi^s$ respectively. A computationally convenient expression for the phase indicator function is given in terms of a signed distance function, $\phi(\mathbf{x}(\mathbf{X}))$, which returns the normal distance away from the interface as a function of material coordinates, and the Heaviside function, H, as

$$\chi = H(\phi(\mathbf{x}(\mathbf{X}))) \tag{21}$$

Morel [44] shows that the gradient of this function represents a signed measure of the surface area per unit volume

$$(\text{grad}\chi)_k = -n_k \delta_S, \tag{22}$$

where $\delta_S$ is a surface Dirac delta distribution which is nonzero along the interface surface, S, and **n** is the unit outward pointing normal to the interface. In other fields (22) is called the interfacial area concentration. It is helpful to create new volume via the intersection of other volumes for constructing the initial volume fraction of the pillar and matrix phases. If we have volume fractions $\chi^1$ and $\chi^2$ representing two different volumes, then the volume created by their intersection is given by [45]

$$\chi^{1 \cap 2} = \chi^{(1)}\chi^{(2)}. \tag{23}$$

When we obtain the volume fraction from the intersection of two volumes like in (23), the interfacial area concentration represents a piecewise surface with two continuous parts obtained from taking the gradient of (23). Using the product rule on (23) gives

$$\mathrm{grad}\chi^{1 \cap 2} = -n_k^1 \delta_{S^1}\chi^{(2)} - n_k^2 \delta_{S^2}\chi^{(1)} \tag{24}$$

where $\delta_{S^1}$ is supported on interface 1 and $\delta_{I^2}$ is supported on interface 2. The first term on the right-hand side in (24) represents the portion of interface 1 contained in volume 2 from the product $\delta_{S^1}\chi^2$. The second term represents the portion of interface 2 contained in volume 1. The last expression can be written as,

$$\mathrm{grad}\chi^{1 \cap 2} = \eta^{1 \cap 2,1} + \eta^{1 \cap 2,2} \tag{25}$$

where $\eta^{1 \cap 2,1} = -n_k^1 \delta_{S^1}\chi^{(2)}$ and $\eta^{1 \cap 2,2} = -n_k^2 \delta_{S^2}\chi^{(1)}$. This last relation illuminates the fact that the surface of the intersection volume has two boundary components represented by $\eta^{1 \cap 2,1}$ and $\eta^{1 \cap 2,2}$. We note that $\eta^{1 \cap 2,1}$ and $\eta^{1 \cap 2,2}$ are not separately the gradient of a scalar, but when summed, they represent the gradient of the phase function $\chi^{1 \cap 2}$. To construct the thin film geometry, we start by showing the relevant volumes that are used to construct $\chi^p, \chi^m$ and $\chi^s$ in Fig. 4

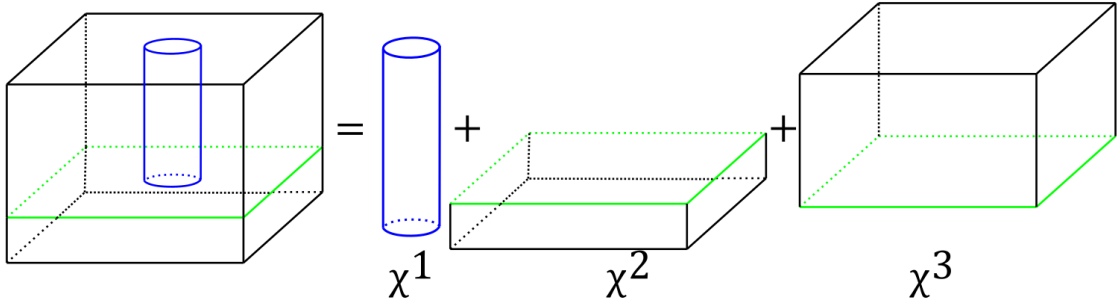

Fig. 4: Depiction of the three different sets of volumes used to create the thin film geometry.

In the notation used in Fig. 4, the pillar volume fraction is given as the intersection of $\chi^{(1)}$ with $\chi^{(3)}$

$$\chi^p = \chi^{(1)}\chi^{(3)}. \tag{26}$$

Next, the matrix phase can be written as the complement of $\chi^p$ in $\chi^{(3)}$

$$\chi^m = \chi^{(3)} - \chi^p \tag{27}$$
$$= \chi^{(3)}(1 - \chi^{(1)}).$$

Lastly, the substrate phase is equal to

$$\chi^s = \chi^{(2)}. \tag{28}$$

From (24), the interfacial area concentration of the pillar and matrix phase consists of two parts,

$$\mathrm{grad}\chi^p \equiv \eta^{pm} + \eta^{ps} \tag{29}$$
$$= \mathrm{grad}(\chi^{(1)})\chi^{(3)} + \chi^{(1)}\mathrm{grad}(\chi^{(3)}).$$

In this expression, $\eta^{pm}$ represents the area concentration for the pillar-matrix interface and $\eta^{ps}$ represents the area concentration for the pillar-substrate interface and may be written in the form,

$$\eta^{pm} = -n_k^{pm}\delta_{S^{pm}} \tag{30}$$

$$\eta^{ps} = -n_k^{ps}\delta_{S^{ps}},$$

where $\delta_{S^{pm}} = \delta_{S^1}H(\phi^{(3)})$ and $\delta_{S^{ps}} = \delta_{S^3}H(\phi^{(1)})$. $\eta^{ms}$ can be constructed similarly by taking the gradient of (27)

$$\text{grad}\chi^m \equiv \eta^{ms} - \eta^{pm} \tag{31}$$
$$= \text{grad}(\chi^{(3)})(1-\chi^{(1)}) - \chi^{(3)}\text{grad}(\chi^{(1)}).$$

and noting that $\eta^{pm} = \chi^{(3)}\text{grad}(\chi^{(1)})$ from (29).

We end this section by giving explicit expressions for each phase. We do this by first constructing $\chi^{(1)}$, which represents the collection of pillars in an infinite domain, using a set of signed distance functions in the form

$$\phi^{(1)} = \sqrt{(x-x_c)^2 + (y-y_c)^2} - R \tag{32}$$

where $x_c$ and $y_c$ denote the x and y coordinate of the pillar's centroid in the x-y plane and $R$ is the radius of the pillar. Both $\chi^{(2)}$ and $\chi^{(3)}$ can be constructed using signed distance functions of the form,

$$\phi = \sqrt{(z-z_c)^2} - z_w/2. \tag{33}$$

This last expression can be viewed as giving the signed distance away from two plates centered at $z_c$ and spaced a distance of $z_w$ apart. It returns a negative distance inside the plates and a positive distance outside the plates. From (21) we can write a smooth representation of $\chi^{(1)}$ by smoothing out the Heaviside function to obtain

$$\chi^{(1)} = H(\phi^{(1)}) \approx \frac{1}{1+\exp\left(-\frac{\phi^{(1)}}{L}\right)} \tag{34}$$

where L is a parameter that controls the interface width. The interested reader is referred to Appendix B, where the relationship between the density fields given in (30) are related to ones written in the reference configuration. Given the expressions (33) and (34) the pillar, matrix and substrate phases can be constructed using (26),(27), and (28).

**3.5 Constitutive equations**

To solve the crystal equilibrium equations, we need to specify the stress in terms of the elastic strain given by a constitutive relationship between the two. This constitutive relationship must satisfy the principle of material frame indifference [46]. An easy way to satisfy the principle of material frame indifference in the multiplicative decomposition framework is to specify the PK2 stress in the microstructure space constitutively. Assuming a linear elastic material in each phase, we write this constitutive relationship as

$$\mathbf{S}^e = \mathbf{C} : \mathbf{E}^e, \tag{35}$$

where $\mathbf{C}$ is the fourth-order elasticity tensor. The elasticity tensor is a piecewise constant function over the domain due to the different phases present in the thin film

$$\mathbf{C} = \chi^m \mathbf{C}^m + \chi^s \mathbf{C}^s + \chi^p \mathbf{C}^p, \tag{36}$$

where $\mathbf{C}^m, \mathbf{C}^s$ and $\mathbf{C}^p$ are the elasticity tensors for the matrix, substrate, and pillar phases, respectively.

We must also constitutively specify the surface stress. To do this, the elasticity effects at the interface associated with the stretching of bonds in each region of the interface are ignored and only include the stress induced by the creation of the surface by constitutively specifying the surface stress as

$$\boldsymbol{\sigma}^{(\Gamma)} = \gamma \mathbf{P}. \tag{37}$$

In the last expression, $\mathbf{P} = \mathbf{I} - \mathbf{n} \otimes \mathbf{n}$ is the projection operator onto the interface, $\mathbf{n}$ is the outward unit normal to the interface, and $\gamma$ is the interface free energy. The interface free energy is a piecewise constant function over the interface because of the different interfaces present in the film,

$$\gamma = \gamma^{pm} \delta_{S^{pm}} + \gamma^{ms} \delta_{S^{ms}} + \gamma^{ps} \delta_{S^{ps}}. \tag{38}$$

Using the last expression, the constitutive specification of the surface stress becomes

$$\boldsymbol{\sigma}^{(\Gamma)} = \gamma^{pm} \mathbf{P}^{pm} \delta_{S^{pm}} + \gamma^{ms} \mathbf{P}^{ms} \delta_{S^{ms}} + \gamma^{ps} \mathbf{P}^{ps} \delta_{S^{ps}} \tag{39}$$

We treat the interface free energy as a coefficient indicating the strength of interface surface tension. Fundamentally, we adopt Fowkes equation to estimate the interface free energy [47], [48]. Fowkes has demonstrated that the interfacial energy between two phases can be obtained by summing up the surface free energies minus the reversible work required to separate them. In other words, the interfacial free energy can be thought of as the energy required to make two phases adhered to each other [22]. Following Fowkes, we estimate the interfacial free energy between phase $\alpha$ and phase $\beta$ as

$$\gamma^{\alpha\beta} = \gamma^{\alpha} + \gamma^{\beta} - W^{\alpha\beta} \tag{40}$$

where $\gamma^{\alpha}$ and $\gamma^{\beta}$ are the surface free energies of phase $\alpha$ and phase $\beta$, respectively. $W^{\alpha\beta}$ is the work of adhesion, a quantity known to be the energy required to separate two bodies $\alpha$ and $\beta$ via cleavage [22]. Fowkes has estimated this energy based on the assumption that London dispersion dipole forces exist between two materials. We assume that at the metal/metal-oxide interface, partial charges of weaker bonding nature are formed due to charge transfer between the two phases, and dipole-dipole forces exist across the interface [49]. In (40), the work of adhesion is estimated based on the assumption that dipole forces exist

$$W^{\alpha\beta}_{dispersion} = 2\sqrt{\gamma_d^\alpha \gamma_d^\beta} \tag{41}$$

Equation (41) is estimated as a geometric mean of surface free energies due to dispersion forces between the two phases in contact. (42) below estimates $\gamma_d^\alpha$ given information about the polarizability $\mathcal{P}_\alpha$, ionization potential $I_\alpha$, the number of representative volume elements per unit area which can be obtained using the material's density and molar mass $N_\alpha$ and $r_\alpha$ is the interplanar distance that is the d-spacing defined in Bragg's equation

$$\gamma_d^\alpha = \frac{-\pi N_\alpha^2 \, \mathcal{P}_\alpha^2 \, I_\alpha}{8 \, r_\alpha^2}. \tag{42}$$

## 4. Numerical solution

It is challenging to solve interface-type problems due to the coupling of the interface boundary conditions in (18). This section derives a weak form defined on the whole domain using the strong form given in each phase in (17)-(19). To derive the weak form, we start by multiplying the stress equilibrium equation (17) by a virtual displacement, $\delta u_i^I$, and integrate over each phase domain indexed by I, to obtain

$$\int_{\Omega^\alpha} \left(\frac{\partial \sigma_{ij}^\alpha}{\partial x_j}\right) \delta u_i^\alpha dv = 0. \tag{43}$$

Integration by parts and Gauss' theorem yields

$$0 = \int_{\partial\Omega^\alpha} \delta u_i^\alpha \sigma_{ij}^\alpha n_j^\alpha \, ds - \int_{\Omega^\alpha} \frac{\partial(\delta u_i^\alpha)}{\partial x_j} \sigma_{ij}^\alpha dv. \tag{44}$$

Summing over each phase domain in (44), noting that $\delta u_i^\alpha$ is continuous across the interface, and realizing that each of the phase boundaries, $\partial\Omega^\alpha$, either lies on a subset of the interface $\Gamma$, or on a subset of the surface of the body $\partial\Omega$ giving

$$0 = \int_{\partial\Omega} \delta u_i \sigma_{ij} n_j \, ds + \int_\Gamma \delta u_i [\sigma_{ij} n_j] \, ds - \int_\Omega \frac{\partial(\delta u_i)}{\partial x_j} \sigma_{ij} dv, \tag{45}$$

where $\sum_\alpha \int_{\partial\Omega} \delta u_i^\alpha \sigma_{ij}^\alpha n_j^\alpha \, ds = \int_{\partial\Omega} \delta u_i \sigma_{ij} n_j \, ds$ and $\sum_\alpha \int_{\Omega^\alpha} \frac{\partial(\delta u_i^\alpha)}{\partial x_j} \sigma_{ij}^\alpha dv = \int_\Omega \frac{\partial(\delta u_i)}{\partial x_j} \sigma_{ij} dv$ have been used. In the previous expression, the normal is chosen to point into the phase with the larger index where the index ordering comes from the set [p, s, m]. From the interface balance (18) and the constitutive relation for the surface stress (39), the weak form becomes

$$0 = \int_{\partial\Omega} \delta u_i \sigma_{ij} n_j \, ds - \sum_{\alpha,\beta>\alpha} \int_\Gamma \delta u_i \left(\gamma^{\alpha\beta} 2\kappa n_i^{\alpha\beta} \delta_{S^{\alpha\beta}} + \gamma^{\alpha\beta} \epsilon_{jki} n_j^{\alpha\beta} \delta_C T_k^{\alpha\beta}\right) ds$$
$$- \int_\Omega \frac{\partial(\delta u_i)}{\partial x_j} \sigma_{ij} dv, \tag{46}$$

where $\gamma^{\alpha\beta}$ has been assumed to be constant and the surface divergence of the projection operator is given by $\text{div}_\Gamma(\mathbf{P}) = 2\kappa\mathbf{n}$ where $\kappa$ is the mean curvature. In this expression, $\delta_C$ represents a Dirac delta distribution concentrated at the common line, and $\mathbf{T}^{\alpha\beta}$ is the tangent of the common line oriented consistently with the boundary of the $\Gamma^{\alpha\beta}$ interface. We give details of the derivation for the terms in the integral over the interfaces in Appendix A. The summation of terms containing $\gamma^{\alpha\beta}\epsilon_{jki}n_j^{\alpha\beta}\delta_C T_k$ vanish from the common line balance in (19) and noting that $\mathbf{v}^{\alpha\beta} = \mathbf{n}^{\alpha\beta} \times \mathbf{T}^{\alpha\beta}$.

From (30) and the compact support of $\delta_S\mathrm{IJ}$ we can obtain the final expression of the weak form as

$$0 = \int_{\partial\Omega} \delta u_i \sigma_{ij} n_j \, ds - \sum_{\alpha,\beta>\alpha} \int_\Omega \delta u_i \gamma^{\alpha\beta} 2\kappa\eta_i^{\alpha\beta} \, dv - \int_\Omega \frac{\partial(\delta u_i)}{\partial x_j} \sigma_{ij} dv. \tag{47}$$

This expression is the final form for the weak form of the momentum balance. It contains terms (the middle terms on the right-hand side of (47)) that can be viewed as concentrated body forces due to the curvature of the interface. In addition to the material parameters, $\gamma^{\alpha\beta}$ and $\mathbf{C}^m$, the area concentrations $\mathbf{\eta}^{\alpha\beta}$ are needed to obtain a solution to (47). The numerical solution to (47) has been implemented in a Multiphysics object oriented open-source finite element framework, MOOSE [50] using first order shape functions on hexahedron elements. It implemented using a Total Lagrange solution method.

## 5. Energetics of the elastic model for VAN thin film

This section aims to identify a generic way of comparing the total energy of various equivalent pillar configurations. To establish what is meant by equivalent pillar configurations, we introduce a representative area element (RAE) of the film and a representative volume element (RVE) equal to the height of the film times the RAE. Next, we use this RAE to develop specific average quantities that are used to parametrize the system's total energy. Specifically, we parametrize the total energy by the total pillar area fraction, the area of the RAE, and the pillar density. Then, two pillar configurations are equivalent if they share the same area fraction and pillar density for a given RAE area. With this equivalence of film morphology established, we can meaningfully compare different pillar morphologies and thus search for a meaningful lowest energy morphology within a set of equivalent RVE's.

We start by introducing a representative volume element. This RVE consists of the substrate and thin film. Fig. 5 shows a typical RVE for (a) hexagon and (b) square pillar configurations where the red color denoted the pillar locations, and the blue corresponds to the matrix phase. The RVE is constructed using a periodic arrangement of pillar locations which can be seen in the figure from the locations of the cut pillars along the boundary of the RVE. In this configuration, there are 4 full pillars contained within the RVE.

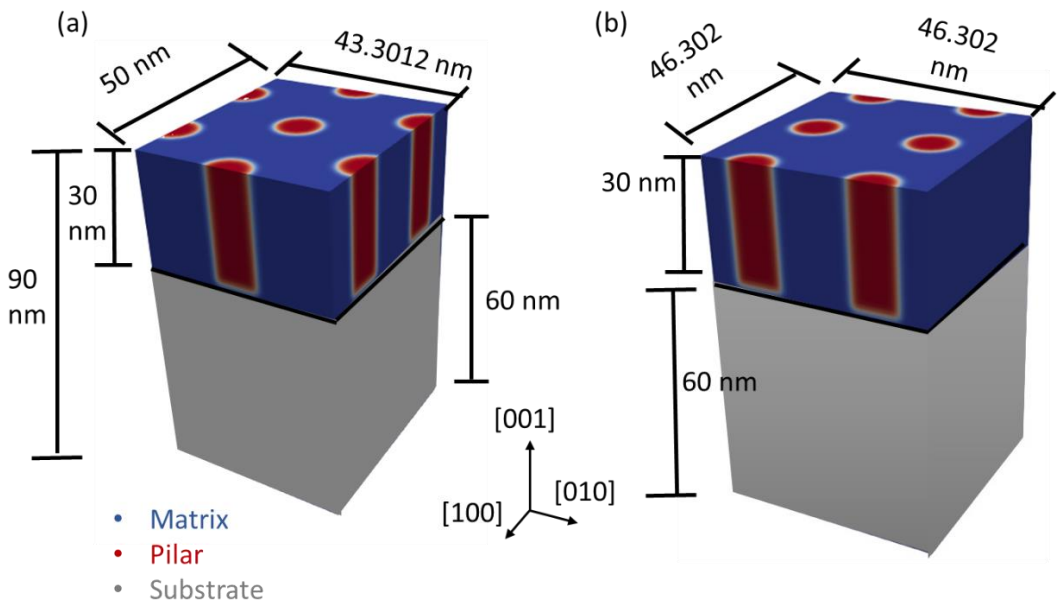

Fig. 5: Size of the RVE for both the (a) hexagonal and (b) square VAN thin film configurations. These dimensions are typical of LSMO-Au system VAN on STO substrate. Both configurations contain an ordered pillar lattice and share both the same pillar density and volume.

When comparing various pillar morphologies, the geometry of the substrate is unchanged, and thus we would like to only analyze changes due to morphological changes of the pillars. Assuming the pillar configurations do not vary throughout the height of the film, we introduce a representative area element (RAE) over the cross-section of the thin film. This RAE is used to compare the various pillar configurations consistently. The RAE is characterized by the total area fraction of the pillar phase and the number of pillars. The total pillar area fraction can be obtained from integrating the pillar volume fraction $\chi^p$ over the RAE at a specific pillar height

$$f = \frac{1}{A_{RAE}} \int_{A_{RAE}} \chi^p dA = \frac{1}{A_{RAE}} \sum_i \int_{A_{p_i}} 1 dA \qquad (48)$$

where $A_{p_i}$ is the cross-sectional area of each pillar. When the pillars all have the same circular cross-section, the total area fraction can be written as

$$f = (N_p \pi R_p^2)/A_{RAE} \qquad (49)$$

where $N_p$ is the number of pillars and $R_p$ is the radius of each pillar. Also, the total area of the RAE is

$$A_{RAE} = N_p \pi R_p^2 + A_m \qquad (50)$$

where $A_m$ is the cross sectional area of the matrix phase contained in the RAE.

We would now like to compare the energetics of each of the pillar configurations. With this in mind, we write the total energy $\mathcal{H}$, which is written as a sum of elastic energy, $\mathcal{U}$, and interfacial energy, $\mathcal{G}$, in terms of the average variables $A_{RAE}$, $n_p$ and $f$,

$$\mathcal{H} = \mathcal{U}(A_{RAE}, n_p, f) + \mathcal{G}(A_{RAE}, n_p, f), \qquad (51)$$

where $n_p$ is the pillar density defined by $n_p = N_p/A_{RAE}$. The total elastic strain energy takes on the form:

$$\mathcal{U}(A_{RAE}, n_p, f) = \int_{\Omega^p} W(\mathbf{E}^e, A_{RAE}, f)\, dV + \int_{\Omega^m} W(\mathbf{E}^e, A_{RAE})\, dV \qquad (52)$$

$$+ \int_{\Omega^s} W(\mathbf{E}^e, A_{RAE})\, dV,$$

where the elastic energy density $W(\mathbf{E}^e, A_{RAE}) = (J^v/2)\mathbf{S}^e : \mathbf{E}^e$ is given by (13) after pulling it to the reference configuration. The first term computes the elastic energy stored in all pillars, and the second and third considers the energy stored in the matrix and substrate phases respectively. Using (35) and (36), (52) can be written as

$$\mathcal{U}(A_{RAE}, n_p, f) = \frac{1}{2}\int_{\Omega} (\mathbf{C} : \mathbf{E}^e) : \mathbf{E}^e J^v\, dV. \qquad (53)$$

In the above, $J^v = \det(\mathbf{F}^v)$ is the determinant of inelastic distortion. On the other hand, the total interfacial energy is computed as follows

$$\mathcal{G}(A_{RAE}, n_p, f) = \int_{\Gamma^{pm}} \gamma^{pm}\, dA + \int_{\Gamma^{ps}} \gamma^{ps}\, dA + \int_{\Gamma^{ms}} \gamma^{ms}\, dA + \int_{\Gamma^{pf}} \gamma^{pf}\, dA + \int_{\Gamma^{mf}} \gamma^{mf}\, dA \qquad (54)$$

In this expression, the first three terms compute the interfacial energy between the pillar/matrix, pillar/substrate, and matrix/substrate, respectively. The last two terms compute the interfacial energy of pillar and matrix respectively at the free surface of the film. Using the surface Dirac delta functions for each interface $\delta_{S^{IJ}}$ the previous expression can be written as

$$\mathcal{G}(A_{RAE}, n_p, f) = \sum_{\alpha,\beta>\alpha} \int_\Omega \gamma^{\alpha\beta} \delta_{S^{\alpha\beta}} \, dV = \sum_{\alpha,\beta>\alpha} \int_\Omega \gamma^{\alpha\beta} |\eta^{\alpha\beta}| \, dV \tag{55}$$

where $\delta_{S^{\alpha\beta}} = |\eta^{\alpha\beta}|$ has been used to obtain the last equality. This last equality highlights the fact that the total energy (51) is dependent on the volume fraction $\chi^\alpha$ of each phase and its gradients through (29).

## 6. Results and discussion

### 6.1 Lattice mismatch deformation simulation

In this section, we provide numerical solutions to the weak form in (47). We apply our model to a computational domain consisting of cylindrical Au pillars in a (LSMO) matrix both of which are placed on a flat (STO) substrate. The material properties of each material used in the simulation are listed below in Table 1.

Table 1. Film/substrate material-type physical properties.

| Material | $C_{11}$ (nN/nm$^2$) | $C_{12}$ (nN/nm$^2$) | $C_{44}$ (nN/nm$^2$) | $\gamma$ (nN·nm)/nm$^2$ | I (eV) | $\mathcal{P}$ (Å$^3$) | $r_\alpha$ (Å) | $N_V$ (Å$^{-3}$) |
|---|---|---|---|---|---|---|---|---|
| Au | 159.1[51] | 136.7[51] | 27.6[51] | 1.627[52] | 9.2257[53] | 8.0[54] | 2.470 | 5.901E+22 |
| LSMO | 228.1[55] | 159.1[55] | 67.1[55] | 0.940[56] | 4.90[57] | 1.521[58] | 1.938 | 1.806E+22 |
| STO | 319[59] | 100[59] | 110[59] | 1.260[60] | 6.90[61] | 3.6469[62] | 1.9525 | 1.581E+22 |

Using the values in Table 1, we simulate a square pillar configuration with 1 pillar in the RVE. This RVE is created using periodic boundary conditions in the plane of the thin film. The box dimensions are $46.53 \times 46.53 \times 90$ nm$^3$ for the square pattern which has an equivalent cross sectional area to the hexagonal configuration which has dimensions of $46.53 \times 50 \times 90$ nm$^3$. In both of these simulation domains the last dimension of 90 (nm) is broken up into a 30 (nm) tall film and 60 (nm) tall substrate. Using the values in Table 1 and equations (40)-(42), the interfacial energies between pillar/matrix, pillar/substrate, and matrix substrate are found to be 2.3290 J/m$^2$, 2.7696 J/m$^2$, and 2.0470 J/m$^2$, respectively. The following simulations are run on 8 node hexahedron elements with a total of $120 \times 120 \times 120$ elements with an interface smoothing parameter of $L = 0.4$ (nm).

Fig. 6 shows the elastic strain energy density which is nonzero due to stress generated from the curved interface and lattice mismatch. The material inside the pillar pushes up against the top free surface, where we observe the pillar phase bulging out of the film. The strain energy is lower in this region due to the stress relaxation effect of the free surface. However, due to the lattice mismatch strain, the bulk of the pillar must be compressed to match with the matrix lattice causing a more substantial development of the strain energy near the middle of the pillar material. This finding agrees with our basic understanding that the pillar should be under compressive strain, whereas the matrix should be under tensile strain due to the lattice parameter of gold (pillar phase) being larger than that of LSMO (matrix phase) [63]. Fig. 6. (a) and (b) show deformation due to lattice mismatch and applied surface tension in the reference and deformed configurations. In the reference configuration, the elastic strain energy density seems to be negligibly higher than in the deformed configuration. In addition to this, we also notice almost ~0 energy density compared to the energies of the bulk materials at the interface, which is expected because we neglected elastic effects at the interface.

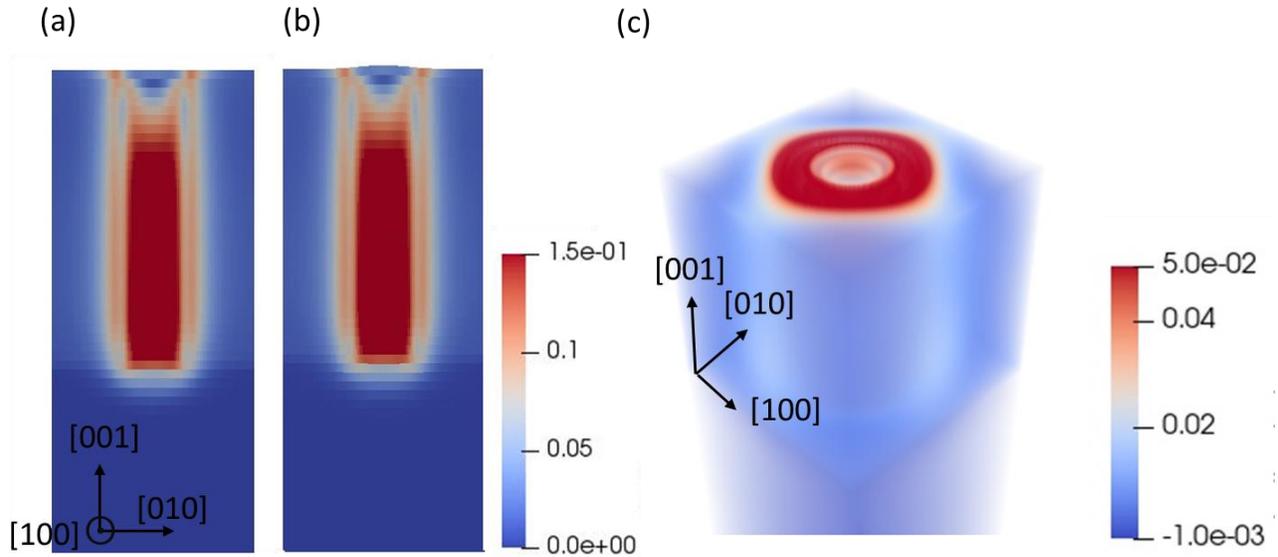

Fig. 6: Elastic strain energy density scaled to units of nN·nm resulting from lattice mismatch deformation (a) in reference configuration, (b) deformed configuration. 3D representation of the elastic strain energy density of the simulation domain. Material simulated is epitaxially deformed single Au pillar embedded in a LSMO matrix that has been grown on $SrTiO_3$ substrate.

Fig. 6 (c) shows a 3D solution to the epitaxial deformation problem. We note that the 3D graphical representation of the strain energy density is showing only the non-zero elements of the final solution. In both reference and deformed configurations, notice that because of residual stresses due to curvature effects and lattice mismatch across the interface, the pillar is pushing down against the substrate, causing a slight dent on the substrate surface. The curvature effect arises due to both the differing lattice constants in each phase and the existence of differing surface tension forces along the common line, causing the pillar's base to shrink. The solution to the derived elastic deformation problem defined in (47) can be extended to study the energetics of microstructure of interest with N number of pillars.

## 6.2 Energy of different pillar morphologies

The VAN microstructure shown in Fig. 1 (a) demonstrates a quasi-random pillar configuration. In addition to studying the energetics of the random configuration, we examine the effects of ordered patterns such as hexagonal and square. We show various pillar configurations with the same area fraction equal to 0.14 below in Fig. 7, which is typical of the STO, LSMO, and Au system [43]. We note that for all configurations shown in Fig. 7, the cross-sectional area for the simulation domain is the same. The box dimensions for the hexagonal and quasi-random configuration are shown in Fig. 5 (a) and for the square box in Fig. 5 (b). The reason for the differing cross-sectional dimensions is because we are forced to make a perfectly square computational domain for the square configuration so that the periodic neighboring pillars maintain the square pillar lattice. When increasing the number of pillars and holding the volume fraction constant, the radii of each pillar must decrease, as is shown in the figure.

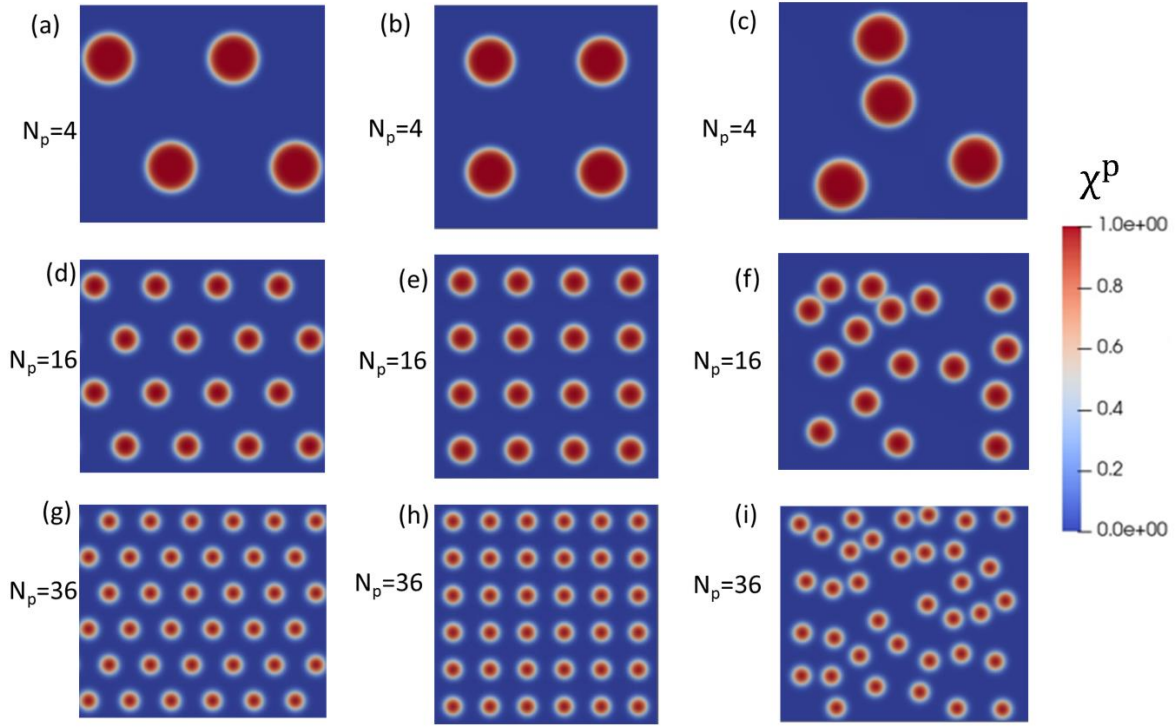

Fig. 7: Depiction of hexagonal pillar configurations in (a), (d), and (g); square pillar configurations in (b), (e) and (h) and quasi-random pillar configurations in (c), (f) and (i) with $N_p = 4, 16, 36$, respectively for the same simulation domain with a volume fraction of 0.14. All figures are scaled with respect to the pillar phase, $\chi^p$.

For each pillar configuration above, the volume fraction is varied by changing the radii of each pillar and the number density of pillars. In Fig. 8, the total energy is plotted, which is calculated using (51), and its components contributed from the elastic strain energy and the total interfacial energy calculated from (53) and (55), respectively. We plot the energy values as a function of pillar density and volume fractions varying around the experimentally measured value of 0.14 [43].

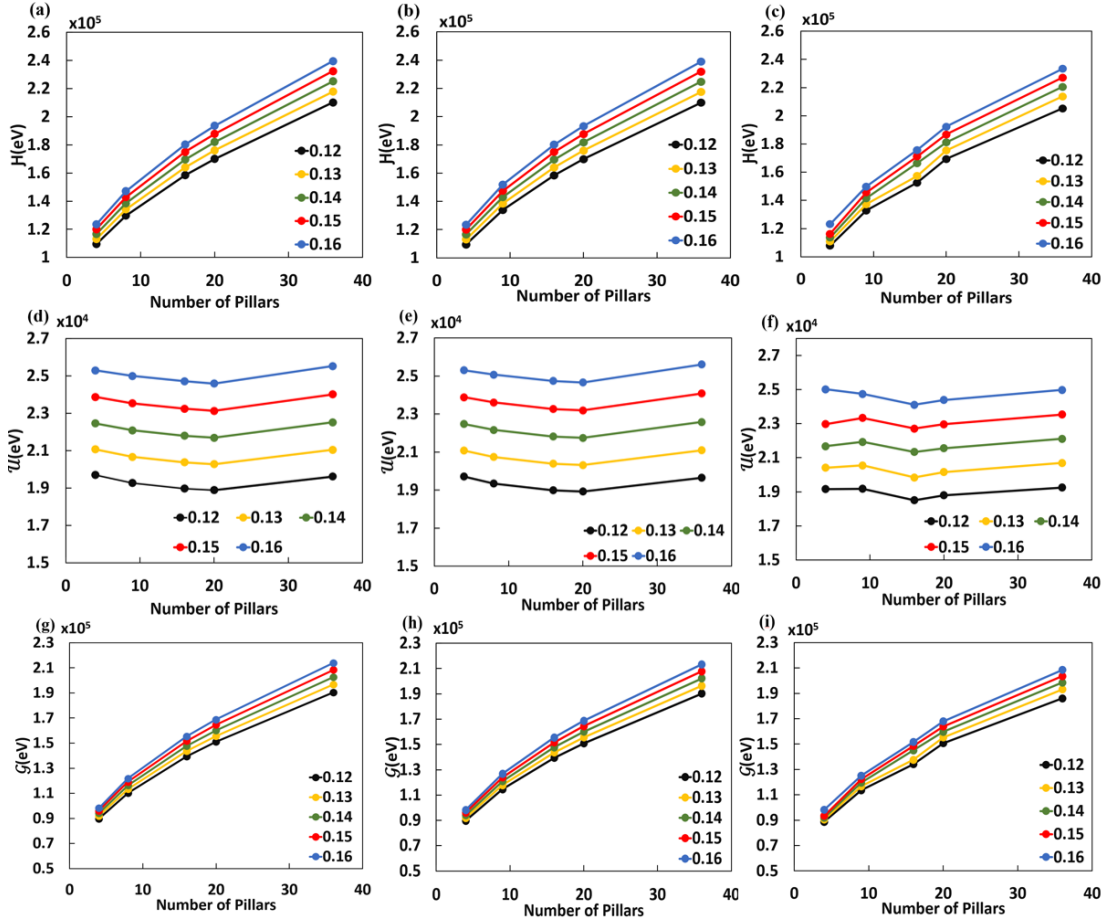

Fig. 8: (a), (b) and (c) are the total energy of hexagonal, square and quasi-random distribution patterns, broken down into their corresponding components (d), (e) and (f) for the total elastic strain energy and (g), (h) and (i) for the total interfacial energy, respectively.

The total energy in Fig. 8 (a), (b) and (c) show a square root functional behavior. When expressing (54) in terms of film geometry and replacing the radius term with $N_p$ and f, it is noted that $\mathcal{G}$ scales with the product $\sqrt{N_p f}$ and has a term linearly in f. At a given volume fraction this scaling suggests that increasing the number of pillars will decrease the radii of the pillars and thus increasing the total area of the matrix. The total energy of each configuration is dominated significantly by the total interfacial energy as seen in (g), (h), and (i). The interfacial energy is highest for the pillar/matrix interface followed by the comparatively lower other interface contributions. The elastic energy shows an interesting functional behavior indicating a minimum value at a particular pillar density. For both square and hexagonal configurations, the critical energy value occurs $N_p = 20$, which corresponds to pillar radii of 2.1946 nm when the volume fraction is 0.14. For the case of the random configuration, the minimum energy occurs at $N_p = 16$ with a radius for each pillar of 2.4556 (nm) for the same volume fraction. We note that these numbers of pillars are in the range found in experiments which can be seen by looking at Fig. 1 (a). In a sample size of 46.53 (nm) × 50 (nm) we obtain an average number of 18 pillars.

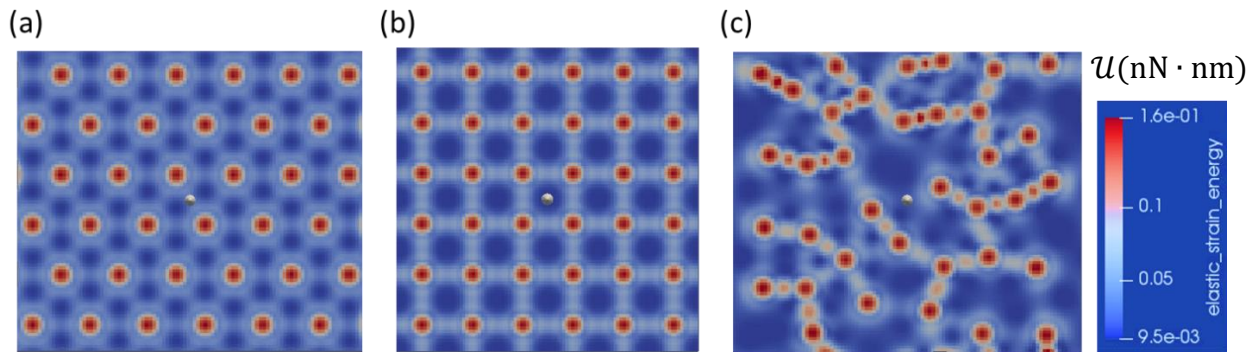

Fig. 9: Elastic energy density of (a) hexagonal (b) square and (c) random configuration for the 36 pillar case. The energetic scale is highlighted to the right and units are in nN·nm

In Fig. 9, we show the elastic strain energy density distribution in the 3 different configurations. The hexagonal configuration has lower total energy than the square. As a result, the elastic energy density between two pillars in the hexagonal becomes slightly smaller than in the square. What's even more interesting is that in Fig. 9, (a) and (b) the non-zero energy density regions draw out the pillar configurational pattern due to the regions of non-zero energy density between nearest neighboring pillars. In both (a) and (b) these non-zero energy density regions form a pattern that is related to the lattice of pillar locations. However, in Fig. 9, (c), no such pattern emerges because of the random locations of each pillar. In (c) we also notice the high elastic energy density between two pillars that are closely spaced. We also notice that in certain areas there are pairs of pillars spaced similarly apart but the regions between the pairs have different elastic energy density, suggesting a shielding effect. Since the total elastic energy is lower for configuration (c) this suggests that the shielding effect is more dominant in (c) than in both (a) and (b).

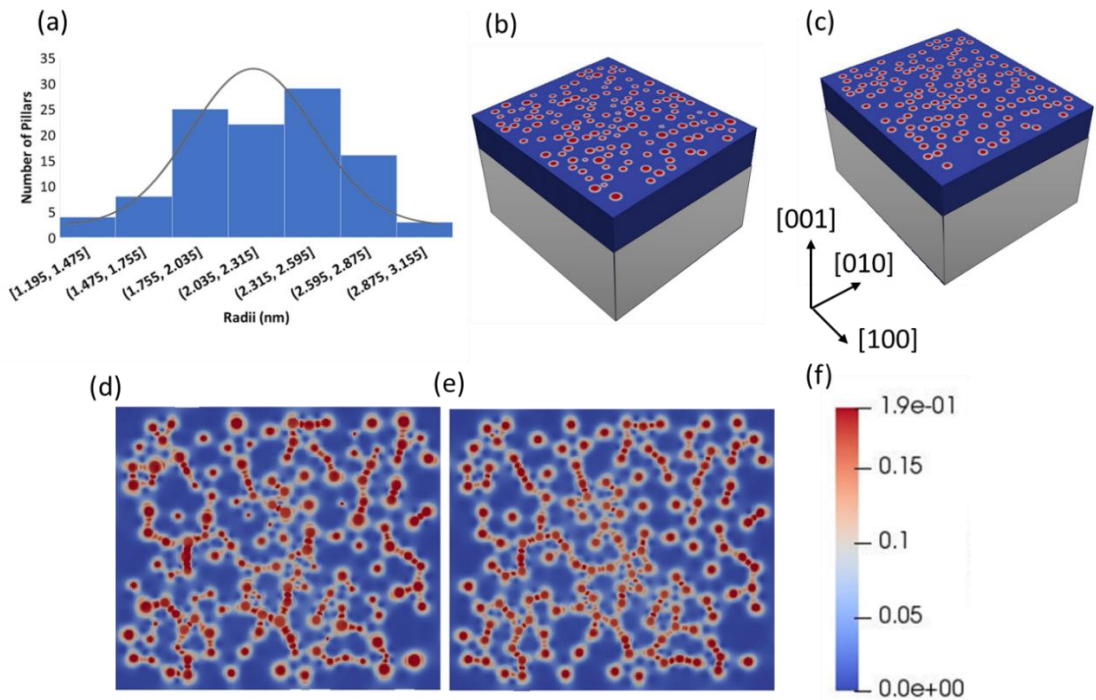

Fig. 10: (a) Normal pillar size distribution with average radius of 2.1232 nm and average volume fraction of 0.14 obtained from Fig. 1 (a). (b) A 144 pillar configuration corresponding to pillar size distribution in (a). (c) A 144 pillar configuration with each pillar radii of 2.1232 nm. (d) and (e) Elastic energy density of xy cross section of (b) and (c), respectively. (f) A scale bar of elastic energy with units of nN·nm

To simulate the microstructure obtained from experiments, as shown in Fig. 1 (a), it would require the pillars in the simulation to vary in size. To vary the pillar size consistent with the experiments, a radii distribution is created to represent the size of the pillars in experiments. To obtain this distribution 5 sample spaces with an area of $150 \times 130$ nm² of Fig. 1 (a) were analyzed. In each of these sample spaces the size of each pillar is measured and the radius distribution is formed using the histogram plot in Fig. 10 (a) and a normal distribution is fit to it. The average area fraction is 0.137 and the average pillar radius is 2.1232 nm. We would like to analyze the effect that having varying pillar sizes has on the total energy of the system. To do this two different configurations were set up where one has a random distribution of pillar sizes and the other has the same radii for each pillar. We note that both configurations have the same cross-sectional area with dimensions 150 x 130 (nm$^{2)}$ and both configurations share the same pillar locations. The configuration with the varying pillar radii is in Fig. 10 (b) and the other in Fig. 10 (c). The box has a height of 90 (nm), of which 30 (nm) is composed of the film. The simulations are run on 8 node hexahedron elements with a total of 300×300×120 elements with an interface smoothing parameter of L = 0.2 (nm). The elastic energy density in (d) and (e) is plotted for both microstructures represented in (b) and (c), respectively. We note that the total elastic and interfacial energies for the configuration with the distribution of pillar radii is $2.0921 \times 10^5$ eV and $1.3136 \times 10^6$ eV while for the average radii configuration is given by $2.0937 \times 10^5$ eV and $1.3317 \times 10^6$ eV, respectively. The total energy including both total elastic and interfacial are in general relatively lower in (d) than (e).

A brief discussion of some important predictions of our model is in order. The pinching effect of the pillar's base in Fig 1 (b) is due to both the presence of capillarity forces and lattice mismatch effects. The developed model demonstrates this effect as can be seen in Fig. 6. In the latter, we also show that the highest elastic energy density is near the curved pillar-matrix interface compared to the relatively low energy density near the substrate interfaces consistent with the more significant lattice mismatch between the pillar-matrix phases. By looking at the same figure, a large stress relaxation effect is observed due to the free surface that leaves a solid cone-like region of low elastic energy density into the bulk of the pillar. The extend of the latter effect may be hard to find by analytic methods alone.

The model also enabled a comparison of the energy of various VAN systems with different pillar morphologies as shown in Fig. 7 for which the line plots of the energy were shown in Fig. 8. Interestingly, the elastic strain energy shows a minimum near the conditions (area fraction, pillar density, and pillar radii) found in experiments [51], giving a validation of the model. For the VAN system studied here, it was found that the interfacial energy was roughly 70% of the total energy and that the pillar-matrix interface was the major contributor to the interfacial energy. In the design of these materials, more attention is paid to engineering the lattice mismatch to minimize the strain energy [20]. In epitaxial engineering, the importance of interfacial energy in controlling the morphology of the film microstructure was reported in previous models [64, 65]. The model further predicted that the random pillar configuration in Fig. 9 has a lower energy which was attributed to an elastic shielding effect, which requires further investigation.

## 6. Conclusion

In this paper, a general 3D elastic model that considers the effects of lattice mismatch and the capillarity forces caused by the curvature nature of the interface at the nanoscale was developed. We derive the boundary value problem of the crystal mechanical equilibrium equations by deriving the bulk, interface, and common line balance laws. This boundary value problem consists of both interface type conditions along with boundary conditions. Furthermore, a weak form for this interface-type boundary value problem is derived by incorporating the interface conditions using a smooth approximation to the sharp interface, and its solution is provided using the finite element method implemented in MOOSE. An RVE was introduced to compare various pillar morphologies meaningfully and simulation results were provided that compare the total energy for various values of area fraction and pillar density. We found that the total energy mainly consists of interfacial energy which may have implications on the design of thin films where constituent materials are typically chosen based on the minimization of lattice mismatch alone. Also it was found that the elastic energy is minimal when the area fraction and the number of pillars is close to what is found in experiments, providing validation of the model. This model can be used to aid in the selection of thin film materials by informing the experimentalist information about how their choice of materials affect the elastic and interfacial energy of the system.

Due to the nature of the interface representation, the model is not limited to solving a deformation-type problem of cylindrical-type interfaces. As a matter of fact, the interface can take any shape. In future work we plan on analyzing the effect that the smoothness of the interface has on the energetics of the system. It would also be possible to extend this model to include lattice mismatch effects coming from compositional changes in each phase along with mismatch effects due to thermal strains. In addition to this elasticity effects at the interface to account for the elastic strain energy present due to the stretching of the interface can be included.

To better understand why a system develops towards a certain morphology will require an understanding of the kinetics of VAN growth. Furthermore, our understanding may be expanded by performing dynamical deformation studies. This can be achieved by implementing the phase field modeling technique to study the interfacial dynamics along with solving the crystal equilibrium equations. In addition to that, a study that involves the understanding of surface diffusion events such as island growth and nucleation of the multiphase VAN during the deposition of the first few monolayers can be conducted using Kinetic Monte Carlo simulations.

**Declaration of competing interests**:

The authors would like to declare that they have no conflicts of competing interest.

**Acknowledgements**:

This work was supported by the U.S. Department of Energy, Office of Science, Basic Energy Sciences (BES) under Award DE-SC0020077. Kyle Starkey acknowledges partial support from the Bilsland Fellowship program at Purdue University.

**Appendix A. Derivation of the common line representation**

The goal of this appendix is to derive the relationship between the open interfaces described by $\boldsymbol{\eta} = -n_k^1 \delta_{I^1} \chi^{(2)} = \text{grad}(\chi^{(1)}) \chi^{(2)}$ and the common line at the boundary of these interfaces. Then we use this relationship to show how the integral in (46) was derived. From the Heaviside representation of $\chi^{(1)}$ and $\chi^{(3)}$

$$\boldsymbol{\eta} = \text{grad}(\chi^{(1)}) \chi^{(3)} = \delta(\phi^{(1)}) \boldsymbol{\nabla}\phi^{(1)} H(\phi^{(2)}). \tag{56}$$

Then using the identity $\text{curl}(u\mathbf{v}) = u\,\text{curl}(\mathbf{v}) + \text{grad}(u) \times \mathbf{v}$, we write,

$$\text{curl}(\boldsymbol{\eta}) = \delta(\phi^{(1)}) H(\phi^{(2)}) \text{curl}(\text{grad}(\phi^{(1)})) \tag{57}$$
$$+ \text{grad}\left(\delta(\phi^{(1)}) H(\phi^{(2)})\right) \times \text{grad}(\phi^{(1)}).$$

The first term is zero because $\text{curl}(\text{grad}(\cdot)) = 0$. Expanding the second term yields

$$\text{curl}(\boldsymbol{\eta}) = \left[\delta(\phi^{(1)})\,\text{grad}\left(H(\phi^{(2)})\right) + \text{grad}\left(\delta(\phi^{(1)})\right)H(\phi^{(2)})\right] \times \text{grad}(\phi^{(1)}) \quad (58)$$

$$= \left[\delta(\phi^{(1)})\delta(\phi^{(2)})\text{grad}(\phi^{(2)}) + \delta'(\phi^{(1)})\text{grad}(\phi^{(1)})H(\phi^{(2)})\right]$$
$$\times \text{grad}(\phi^{(1)})$$

Where the second term vanishes because $\text{grad}(\phi^{(1)}) \times \text{grad}(\phi^{(1)}) = 0$ leaving us with,

$$\text{curl}(\boldsymbol{\eta}) = \delta(\phi^{(1)})\delta(\phi^{(2)})\text{grad}(\phi^{(2)}) \times \text{grad}(\phi^{(1)}) \quad (59)$$

Remembering that the volume $\chi^{(1)}\chi^{(2)}$ represents the intersection of two volumes $\chi^{(1)}$ and $\chi^{(2)}$ and that the boundary of this volume consists of two parts given in (24) where $\delta_{S_{\tilde{c}}^1} = \delta(\phi^{(1)})|\text{grad}(\phi^{(1)})|$, $\delta_{S_{\tilde{c}}^2} = \delta(\phi^{(2)})|\text{grad}(\phi^{(2)})|$, $\mathbf{n}^1 = \frac{\text{grad}(\phi^{(1)})}{|\text{grad}(\phi^{(1)})|}$ and $\mathbf{n}^2 = \frac{\text{grad}(\phi^{(2)})}{|\text{grad}(\phi^{(2)})|}$. With these relations we write (59) as

$$\text{curl}(\boldsymbol{\eta}) = \delta_{S_{\tilde{c}}^1}\delta_{S_{\tilde{c}}^2}(\mathbf{n}^1 \times \mathbf{n}^2) = \delta_C \mathbf{T} \quad (60)$$

where the last equality is due to the intersection of the two closed surfaces $S_{\tilde{c}}^1$ and $S_{\tilde{c}}^2$ which results in a curve, denoted by $C$, with tangent $\mathbf{T} = \mathbf{n}^1 \times \mathbf{n}^2$.

This relation will help derive (46). We start from equation (45) which we copy here below,

$$0 = \int_{\partial\Omega} \delta u_i^\alpha n_j^\alpha \sigma_{ij}^\alpha \, dS + \int_\Gamma \delta u_i [\sigma_{ij} n_j] \, dS - \int_\Omega \frac{\partial(\delta u_i)}{\partial x_j} \sigma_{ij} dV.$$

To derive (46), we must take the surface divergence of the surface stress from the interface balance in (18). We start by writing the surface stress as in (39) and take its surface divergence,

$$\text{div}_\Gamma\left(\boldsymbol{\sigma}^{(\Gamma)}\right) = \sum_{\alpha,\beta>\alpha}\left[\mathbf{P}^{\alpha\beta \text{T}}\text{grad}\left(\gamma^{\alpha\beta}\delta_{S_o^{\alpha\beta}}\right) + \gamma^{\alpha\beta}\delta_{S_o^{\alpha\beta}}\text{div}_\Gamma\left(\mathbf{P}^{\alpha\beta}\right)\right] \quad (61)$$

$$= \sum_{\alpha,\beta>\alpha}\left[\mathbf{P}^{\alpha\beta \text{T}}\text{grad}\left(\gamma^{\alpha\beta}\delta_{S_o^{\alpha\beta}}\right) + \gamma^{\alpha\beta}2\kappa\delta_{S_o^{\alpha\beta}}\mathbf{n}^{\alpha\beta}\right].$$

The second term $\gamma^{\alpha\beta}2\kappa\delta_{S_o^{\alpha\beta}}\mathbf{n}^{\alpha\beta}$ is already in a form consistent with one of the terms in (46). We focus on the first term and note that each of the open surfaces can be written as $\delta_{S_o^{\alpha\beta}} = \delta(\phi^{(1)})|\text{grad}(\phi^{(1)})|H(\phi^{(2)}) = \delta_{S_c^1}H(\phi^{(2)})$. We note that the $\phi^1$ and $\phi^2$ do not necessarily represent the same functions for each interface represented by $\delta_{S_o^{\alpha\beta}}$. We chose these functions such that $n^1 = \text{grad}(\phi^1)/|\text{grad}(\phi^1)|$ shares the normal of the open surface representing the interface. With this representation of $\delta_{S_o^{\alpha\beta}}$, the first term in (61).**Error! Reference source not found.** in the sum becomes

$$\mathbf{P}^{\alpha\beta \text{T}}\text{grad}\left(\gamma^{\alpha\beta}\delta_{S_o^{\alpha\beta}}\right) = \mathbf{P}^{\alpha\beta \text{T}}\gamma^{\alpha\beta}\left(\text{grad}(\delta_{S_c^1})H(\phi^2) + \delta_{S_c^1}\delta_{S_c^2}\mathbf{n}^2\right) \quad (62)$$

$$= \mathbf{P}^{\alpha\beta \text{T}}\gamma^{\alpha\beta}\delta_{S_c^1}\delta_{S_c^2}\mathbf{n}^2$$

$$= \mathbf{n}^{\alpha\beta} \times \left(\mathbf{n}^2 \times \mathbf{n}^{\alpha\beta}\right)\gamma^{\alpha\beta}\delta_{S_c^1}\delta_{S_c^2}$$

$$= \gamma^{\alpha\beta}\mathbf{n}^{\alpha\beta} \times \left(\delta_C \mathbf{T}^{\alpha\beta}\right)$$

where we have used $\mathbf{P}^{\alpha\beta \text{T}}\text{grad}(\delta_{S_c^1}) = \text{grad}_\Gamma(\delta_{S_c^1}) = 0$ and $\mathbf{P}^{\alpha\beta \text{T}}\mathbf{n}^2 = \mathbf{n}^{\alpha\beta} \times \left(\mathbf{n}^2 \times \mathbf{n}^{\alpha\beta}\right)$. With (62), the surface divergence of the surface stress can be written as

$$\text{div}_\Gamma\left(\boldsymbol{\sigma}^{(\Gamma)}\right) = \sum_{\alpha,\beta>\alpha}\left[\gamma^{\alpha\beta}\mathbf{n}^{\alpha\beta} \times \left(\delta_C \mathbf{T}^{\alpha\beta}\right) + \gamma^{\alpha\beta}2\kappa\delta_{S_o^{\alpha\beta}}\mathbf{n}^{\alpha\beta}\right]. \quad (63)$$

This last expression is the one used in writing (46)

**Appendix B. Relationship between deformed and reference area concentrations**

Since the volume fraction is a scalar written in terms of material coordinates, we have the instantaneous relationship

$$\chi = H(\phi(\mathbf{X})) = H(\phi(\mathbf{x}(\mathbf{X}, t))). \tag{64}$$

The referential gradient of the first equality in expression in (64) yields the referential surface area concentration, $\boldsymbol{\eta}_R$,

$$\boldsymbol{\eta}_R = \text{Grad}(\chi) = \text{Grad}\left(H(\phi(\mathbf{X}))\right) \tag{65}$$

$$= \delta(\phi(\mathbf{X}))\text{Grad}(\phi(\mathbf{X})).$$

From the definition of the unit normal $\mathbf{n}_R = \text{Grad}(\phi(\mathbf{X}))/|\text{Grad}(\phi(\mathbf{X}))|$ and the surface Dirac delta distribution $\delta_{S_R} = \delta(\phi(\mathbf{X}))|\text{Grad}(\phi(\mathbf{X}))|$ we write the above expression as

$$\boldsymbol{\eta}_R = \delta_{S_R}\mathbf{n}_R. \tag{66}$$

Taking the spatial gradient of $\chi = H(\phi(\mathbf{x}))$ yields

$$\boldsymbol{\eta} = \text{grad}(\chi) = \text{grad}\left(H(\phi(\mathbf{x}))\right) \tag{67}$$

$$= \delta(\phi(\mathbf{x}))\text{grad}(\phi(\mathbf{x})).$$

From the definition of the unit normal $\mathbf{n} = \text{grad}(\phi(\mathbf{x}))/|\text{grad}(\phi(\mathbf{x}))|$ and the surface Dirac delta distribution $\delta_S = \delta(\phi(\mathbf{x}))|\text{grad}(\phi(\mathbf{x}))|$ we write the above expression as

$$\boldsymbol{\eta} = \delta_S \mathbf{n} \tag{68}$$

Given that $\phi(\mathbf{X}) = \phi(\mathbf{x}(\mathbf{X}, t))$ we find that

$$\text{Grad}(\phi) = \mathbf{F}^T \text{grad}(\phi). \tag{69}$$

With the last expression, we can form a relationship between $\boldsymbol{\eta}$ and $\boldsymbol{\eta}_R$

$$\boldsymbol{\eta}_R = \mathbf{F}^T \boldsymbol{\eta}. \tag{70}$$